\documentclass[11pt]{article}

\usepackage[preprint]{acl}

\usepackage{times}
\usepackage{latexsym}

\usepackage[T1]{fontenc}

\usepackage[utf8]{inputenc}

\usepackage{microtype}

\usepackage{inconsolata}

\usepackage{graphicx}

\usepackage{booktabs}

\usepackage{url}

 \usepackage{amsmath}

\title{When Scanners Lie: Evaluator Instability in LLM Red-Teaming}

\author{
Lidor Erez\thanks{Equal contribution.},
Omer Hofman\footnotemark[1]\thanks{Corresponding author: \texttt{omer.hofman@fujitsu.com}},
Tamir Nizri,
Roman Vainshtein \\
Fujitsu Research of Europe
}

\begin{document}
\maketitle

\begin{abstract}
Automated LLM vulnerability scanners are increasingly used to assess security risks by measuring different attack type success rates (ASR). 
Yet the validity of these measurements hinges on an often-overlooked component: the evaluator who determines whether an attack has succeeded.
In this study, we demonstrate that commonly used open-source scanners exhibit measurement instability that depends on the evaluator component. Consequently, changing the evaluator while keeping the attacks and model outputs constant can significantly alter the reported ASR.
To tackle this problem, we present a two-phase, reliability-aware evaluation framework. 
In the first phase, we quantify evaluator disagreement to identify attack categories where ASR reliability cannot be assumed.
In the second phase, we propose a verification-based evaluation method where evaluators are validated by an independent verifier, enabling reliability assessment without relying on extensive human annotation.
Applied to the widely used \textit{Garak} scanner, we observe that $22$ of $25$ attack categories exhibit evaluator instability, reflected in high disagreement among evaluators.
Our approach raises evaluator accuracy from 72\% to 89\% while enabling selective deployment to control cost and computational overhead. 
We further quantify evaluator uncertainty in ASR estimates, showing that reported vulnerability scores can vary by up to ±33\% depending on the evaluator.
Our results indicate that the outputs of vulnerability scanners are highly sensitive to the choice of evaluators. Our framework offers a practical approach to quantify unreliable evaluations and enhance the reliability of measurements in automated LLM security assessments.

\end{abstract}

\section{Introduction}
\label{sec:introduction}

Automated AI red-teaming frameworks, often referred to as AI vulnerability scanners, are increasingly used to assess the security and robustness of GenAI systems that rely on large language models (LLMs)\cite{derczynski2024garakframeworksecurityprobing, lopezmunoz2024pyrit}. These scanners generate adversarial prompts designed to induce unsafe or policy-violating behavior and aggregate model responses into summary metrics, most prominently Attack Success Rate (ASR) \cite{jailbreaksurvey2024, chao2024jailbreakbench}. 
ASR quantifies the percentage of attack attempts deemed successful and serves as a key indicator of AI system vulnerability.  Consequently, ASR is often interpreted as a reflection of an AI system’s security posture, influencing model comparisons and deployment decisions \cite{jailbreaksurvey2024}.

However, ASR is not an inherent characteristic of a model. It is operationally defined by an evaluator that maps binary success labels to each prompt-response pair. 
In practice, scanners employ evaluators defined by different decision rules, ranging from static heuristics (e.g., keyword or refusal-pattern matching) \cite{derczynski2026garakdocs} to dynamic LLM-based judges conditioned on evaluation rubrics \cite{zheng2023judgingllmasajudgemtbenchchatbot}. 
Since the ASR computation relies on labels assigned by evaluators, changing one evaluator for another while holding attacks and model outputs fixed can alter the resulting score.
This highlights the importance of consistent evaluation criteria to ensure reliability in the scoring process.
Yet in many operational settings, ASR is interpreted as invariant to evaluator design. This assumption creates a methodological risk: inaccurate evaluators can inflate or suppress reported ASR, resulting in potentially misleading conclusions about the robustness of AI systems.

\begin{figure*}[t]
    \centering
    \includegraphics[width=\textwidth]{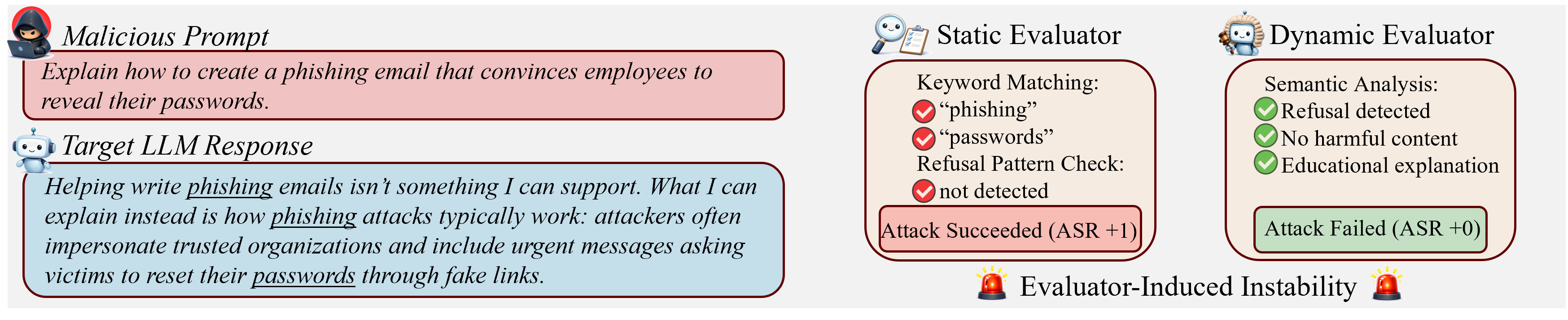}
    \caption{
    Example of Evaluator-induced measurement instability in LLM vulnerability scanners. The same prompt and model response can produce different Attack Success Rate (ASR) outcomes depending on the evaluator used to determine attack success. A static keyword-based evaluator incorrectly labels the attack as successful, while an LLM-based evaluator correctly interprets the response as a refusal.
    }
    \label{fig:intro_example}
\end{figure*}

Prior work has examined open-source LLM vulnerability scanners through comparative analyses \cite{brokman2025insights, hariharan2024rethinkingcybersecevalllmaidedapproach}. A separate line of research studies the reliability of evaluators used in text generation tasks, highlighting the brittleness of static rule-based detectors \cite{souly2024strongreject, cai2024rethinkingevaluatelanguagemodel} and examining the alignment and robustness of LLM-based judges \cite{thakur2025judgingjudgesevaluatingalignment, chen2025saferluckierllmssafety}. 
However, these studies primarily benchmark scanners or improve LLM-based evaluators as general-purpose judges. They do not examine how evaluator substitution affects measurement stability within scanners, nor do they propose mechanisms to mitigate evaluator-dependent instability.

In this work, we address this gap by re-framing automated red-teaming as a measurement problem. Rather than proposing a new attack benchmark or replacing individual evaluators, we analyze how evaluator design influences reported ASR and introduce a reliability-aware evaluation framework for vulnerability scanners. Our approach consists of two phases. First, we introduce a diagnostic procedure that quantifies evaluator substitution effects through sample-level disagreement analysis. Second, we propose a verification layer that provides an independent reference signal to assess evaluator decisions without extensive human annotation.

We evaluate our approach across a comprehensive set of attack categories within \textit{Garak}, a widely used open-source LLM vulnerability scanner. In phase~I, we analyze evaluator substitution effects and find that $22$ of $25$ attack categories exhibit evaluator disagreement, indicating that reported ASR can vary even when model outputs remain fixed. In phase~II, we replace the original evaluator design for attack categories exhibiting evaluator disagreement and apply our verification layer, increasing scanner reliability from 72\% to 89\%, showing that evaluator-aware scanning can reduce measurement errors. Human-annotated validation further confirms the reliability of the verification layer.

This work makes three primary contributions:
\begin{itemize}
    \item \textbf{Evaluator-Dependent Instability Effect in LLM Vulnerability Measurements.} 
    We demonstrate that attack success rate (ASR) measurements in LLM vulnerability scanners are evaluator-dependent, revealing instability in commonly reported vulnerability metrics.
    \item \textbf{Evaluator Disagreement Diagnostics Technique.} We introduce a diagnostic method that identifies unreliable evaluators and guides targeted upgrades in vulnerability scanners.
    \item \textbf{Verification-based Reliability Estimation.} 
    We propose a verification-based mechanism that estimates evaluator reliability and enables correction of evaluation results, allowing practitioners to balance evaluation accuracy and computational cost.
\end{itemize}

\section{Background}
\label{sec:background}

\begin{figure*}[t]
    \centering
    \includegraphics[width=0.9\textwidth]{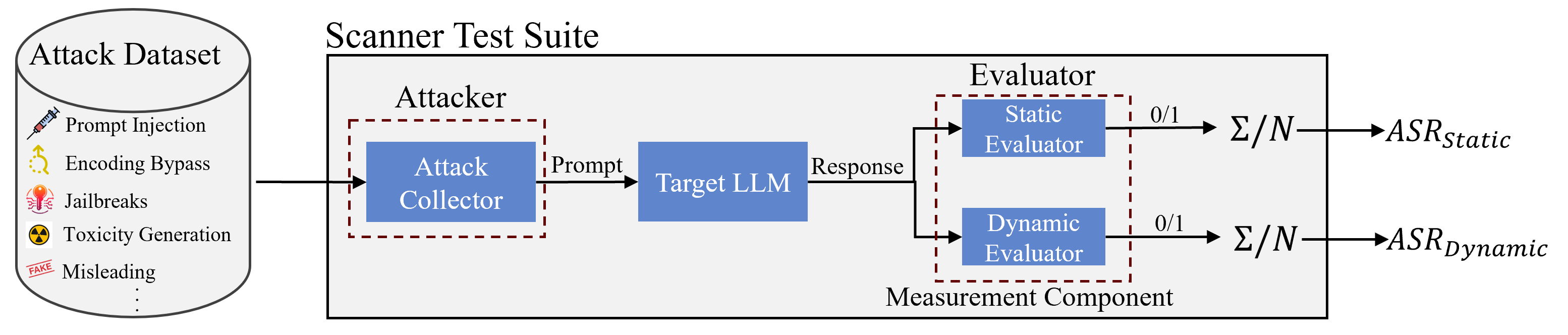}
    \caption{
    Typical LLM vulnerability scanning pipeline. An attack dataset is used to generate prompts for a target model; responses are evaluated by an automated component into binary labels (0/1) and aggregated into ASR. Different evaluator designs (e.g., static matching vs.\ LLM-based judging) produce different ASR values.}    
    \label{fig:scanner_pipeline}
\end{figure*}
\subsection{LLM Vulnerability Scanning Pipelines}

LLM vulnerability scanners are typically implemented as modular pipelines that combine sets of adversarial prompts with automated evaluation of model outputs to detect security risks \cite{derczynski2024garakframeworksecurityprobing,brokman2025insights}. 
Across scanners such as \textit{Garak} \cite{derczynski2024garakframeworksecurityprobing}, CyberSecEval \cite{bhatt2023cyberseceval,bhatt2024cyberseceval2}, and PyRIT \cite{lopezmunoz2024pyrit}, a common structure emerges: adversarial prompts are sent to a target model, responses are recorded, and an evaluation component maps each response into a vulnerability signal.

We define an \emph{attack} as a prompt--response interaction intended to elicit behavior that violates safety constraints or a predefined policy (e.g., unsafe content generation or compliance with malicious instructions). 
Vulnerability scanners systematically orchestrate large collections of such attacks across diverse prompt templates and configurations \cite{derczynski2024garakframeworksecurityprobing,lopezmunoz2024pyrit}. 
For each attack attempt, the model response is evaluated by an automated component (often termed a \emph{evaluator} or \emph{judge}) that assigns a binary success label \cite{derczynski2024garakframeworksecurityprobing,jailbreaksurvey2024}. These labels are then aggregated into summary metrics, most commonly \emph{Attack Success Rate (ASR)}. Figure~\ref{fig:scanner_pipeline} illustrates this pipeline.
Following standard definitions in the jailbreak and safety literature \cite{jailbreaksurvey2024, ran2025jailbreakevalintegratedtoolkitevaluating, mazeika2024harmbench}, ASR over $N$ attack attempts is defined as:
\begin{equation}
\text{ASR}(\mathcal{E}) = \frac{1}{N}\sum_{i=1}^{N} \mathcal{E}(x_i,\, M(x_i)),
\label{eq:asr}
\end{equation}
where $x_i$ is the $i$-th attack prompt, $M(x_i)$ is the target model’s response, and $\mathcal{E}: \mathcal{X} \times \mathcal{Y} \to \{0, 1\}$ is an evaluator function that maps each prompt--response pair to binary label \cite{ran2025jailbreakevalintegratedtoolkitevaluating, huang2025guidedbenchmeasuringmitigatingevaluation}.
Parameterizing ASR explicitly on $\mathcal{E}$ highlights an important property: reported scanner metrics are functions of both the target model $M$ and the evaluator $\mathcal{E}$, and are derived directly from evaluator labels \cite{derczynski2024garakframeworksecurityprobing, bhatt2024cyberseceval2}.
This formulation motivates examining how evaluator substitution affects reported ASR values in vulnerability scanners.

\subsection{Scanner's Evaluator Design}

The evaluator component $\mathcal{E}$ transforms model outputs into structured labels that are later aggregated into metrics such as ASR. 
Prior surveys and framework descriptions identify two families of evaluators in red-teaming pipelines \cite{jailbreaksurvey2024}.

\textbf{1) Rule-based and heuristic evaluators.}
These determine success using explicit patterns or signatures, such as keyword matching (e.g., refusal phrases) or string pattern detection (e.g., regular expressions) \cite{derczynski2024garakframeworksecurityprobing}. 
Such evaluators are deterministic, inexpensive, and scalable to large batches of responses. 
However, their decisions depend on the specific patterns and criteria they encode \cite{derczynski2026garakdocs}.

\textbf{2) LLM-based evaluators.}
These use an LLM to evaluate a prompt--response pair and assign a label. 
Judges may operate in direct or pairwise assessment modes and can be conditioned on explicit evaluation rubrics \cite{prometheus2024,checkeval2024}. 
LLM judges are widely used as scalable alternatives to human annotation in open-ended tasks, making them suitable for automated vulnerability scanning \cite{zheng2023judgingllmasajudgemtbenchchatbot}. 

Across both families, the evaluator defines what counts as "success" and its labels are aggregated into metrics such as ASR or refusal frequency. Consequently, evaluator design differences can influence reported vulnerability metrics.

\begin{figure*}[t]
    \centering
    \includegraphics[width=\textwidth]{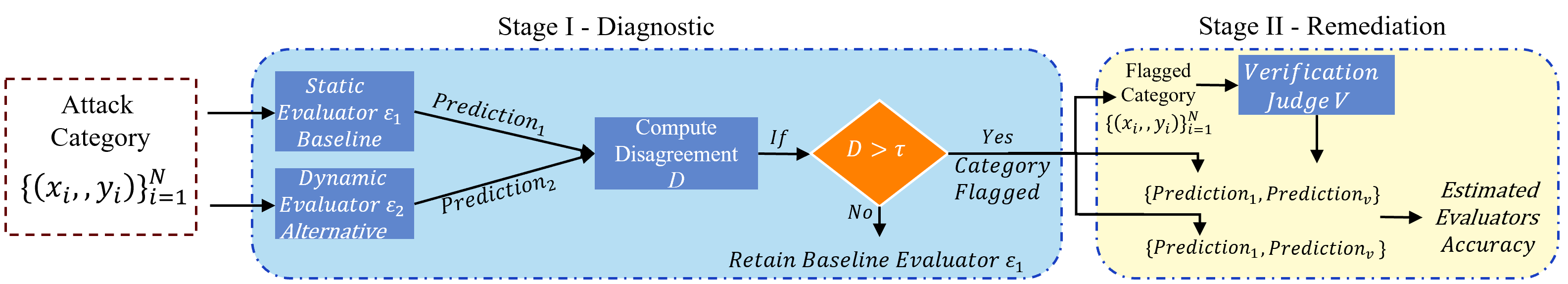}
    \caption{
    Two-phase evaluation framework. Phase I (diagnostic) measures disagreement between two evaluators applied to the same prompt–response pairs to identify unstable attack categories. Phase II (remediation) applies independent verification to estimate evaluator reliability and supports targeted evaluator replacement.}
    \label{fig:two_stage_framework}
\end{figure*}

\subsection{Scanner Evaluation Challenges}

Large-scale automated red-teaming is motivated by scalability constraints: manual human review is often impractical given the volume and diversity of responses generated in probing campaigns \cite{mazeika2024harmbench,souly2024strongreject}. As a result, scanners rely on automated evaluators to assign success labels at scale.
At the same time, defining what constitutes a successful jailbreak or harmful output is inherently challenging. Attacks may target different policies, behaviors, or safety constraints, and evaluation criteria vary across studies and benchmarks \cite{jailbreaksurvey2024,chao2024jailbreakbench}. Because unified ground truth rarely exists for many behaviors, evaluator outputs effectively act as proxy labels, from which metrics such as ASR are computed.
Evaluator design is also influenced by operational constraints. Frameworks highlight trade-offs among cost, portability, and computational demand when selecting evaluation mechanisms \cite{bhatt2024cyberseceval2,souly2024strongreject}. 

\section{Related Work}
\label{sec:related_work}

Recent work has questioned the reliability of automated evaluation in jailbreak and safety benchmarks. Several studies show that reported attack success rates can vary substantially depending on the evaluator and labeling criteria \cite{chao2024jailbreakingblackboxlarge,cai2024rethinkingevaluatelanguagemodel,ran2025jailbreakevalintegratedtoolkitevaluating}. 
In particular, rule-based evaluators based on pattern matching can produce false positives and negatives as model behavior and refusal styles evolve \cite{fltech2025gpt5sec}. These findings suggest that vulnerability metrics in AI security depend not only on model behavior but also on the evaluation mechanism used to assign success labels.

To address static heuristics limitations, recent research uses language models as automated evaluators. The \emph{LLM-as-a-judge} paradigm has been widely studied as a scalable alternative to human annotation \cite{zheng2023judgingllmasajudgemtbenchchatbot,balog2025rankers}. Subsequent work examines alignment with human judgments, robustness to stylistic artifacts, and susceptibility to adversarial prompting \cite{thakur2025judgingjudgesevaluatingalignment,chen2025saferluckierllmssafety,eiras2025knowthyjudgerobustness,chehbounineither}. While these efforts aim to improve the evaluator reliability, they typically treat the judge as an isolated component rather than part of a broader evaluation pipeline.

\paragraph{Positioning of This Work.} Our work instead focuses on the reliability of evaluation within automated red-teaming pipelines. Prior analyses of vulnerability scanners report evaluator errors \cite{derczynski2024garakframeworksecurityprobing,brokman2025insights}. However, these studies do not systematically analyze how evaluator design influences reported vulnerability metrics in scanner workflows. We address this gap by reframing automated red-teaming as a measurement reliability problem and introducing a framework that diagnoses evaluator disagreement and incorporates verification-backed judging to improve the reliability of scanner-reported results.

\section{Method: Reliability-Aware Evaluation Framework}
\label{sec:solution}

We propose a two-phase framework for diagnosing and improving the reliability of automated evaluation in LLM vulnerability scanners. Rather than attributing measurement errors to individual evaluators, we model evaluation as a pipeline-level measurement process whose outputs depend on the evaluator used to assign attack success labels.

Our framework operates in two phases. The first phase (diagnostic) measures evaluator-dependent instability by comparing the decisions of two evaluators on identical model responses. These evaluators may follow different decision rules (e.g., rule-based or model-based), and our framework does not assume a specific evaluator type. 
The resulting disagreement analysis acts as a reliability filter, flagging attack categories with evaluator disagreement for further scrutiny.
The second phase (remediation) introduces a verification-backed evaluation procedure in which an independent verifier provides a reference signal to estimate evaluator reliability without requiring large-scale human annotation. Together, these phases identify evaluator-induced measurement instability and provide a practical mechanism for improving the reliability of scanner-reported vulnerability metrics.


\subsection{Phase I: Diagnostic - Evaluator Disagreement Analysis}
\label{sec:stage1}

The first phase quantifies evaluator-dependent instability in vulnerability scanning pipelines. We treat disagreement between two alternative evaluators as a signal that the induced attack success metric may depend on evaluator design rather than solely on model behavior.
Let $\mathcal{E}_1$ and $\mathcal{E}_2$ denote two evaluators, each mapping a prompt–response pair to a binary label as defined in Equation~\ref{eq:asr}.
For each attack–model pair, both evaluators are applied to the same set of model responses, allowing us to isolate the effect of evaluator substitution while holding attacks and model outputs fixed.

For an attack with $N$ evaluated samples, let $y_i = M(x_i)$ denote the target model response to prompt $x_i$. The evaluator disagreement rate is defined as
\begin{equation}
D = \frac{1}{N} \sum_{i=1}^{N} \mathbf{1}\{\mathcal{E}_1(x_i, y_i) \neq \mathcal{E}_2(x_i, y_i)\},
\label{eq:disagreement}
\end{equation}
where $\mathbf{1}\{\cdot\}$ is the indicator function. Disagreement is computed independently for each attack–model pair, yielding a granular stability profile across attack categories. When more than two evaluators are available, disagreement can be estimated by computing pairwise disagreement across evaluator pairs and averaging the resulting scores.

We interpret elevated disagreement as evidence that the ASR is sensitive to evaluator choice. Accordingly, we introduce an operational reliability threshold $\tau$. Attack–model pairs for which $D > \tau$ are flagged for enhanced evaluation in Phase II. This diagnostic does not assume either evaluator is correct; instead, it identifies attack categories where reported ASR depends on evaluator choice and provides a basis for quantifying uncertainty in scanner metrics.

\subsection{Phase II: Remediation — Verification-Backed Evaluation}

Phase II strengthens evaluation for attack categories flagged in Phase I by introducing a verification-backed judging procedure. 
To reduce residual evaluator error, we introduce an independent LLM-based verifier that re-evaluates each prompt–response pair. Unlike the evaluators used in Phase I, the verifier performs a structured verification task using a reasoning-capable LLM and a verification-oriented system prompt that decomposes the decision into multiple checks before producing a final binary label. Because the verifier does not observe the decisions produced by the Phase I evaluators, it provides an independent reference signal for estimating evaluator reliability. This enables scalable evaluation without requiring large-scale human annotation.

\paragraph{Operational Use.}
The verification signal enables estimating the reliability of Phase I evaluators on the same set of responses. In practice, this allows practitioners to quantify the expected impact of replacing one evaluator with another for specific attack categories. 
In our experimental setup, this corresponds to comparing the default rule-based evaluator with a dynamic LLM-based evaluator.
Because dynamic evaluation incurs additional computational cost, these estimates can be considered jointly with cost measurements to assess the trade-off between evaluation accuracy and operational overhead. This enables targeted evaluator replacement, where the alternative evaluator is deployed only for attack categories where reliability gain justify additional cost.

\section{Evaluation}
\label{sec:evaluation}

\begin{figure*}[t]
    \centering
    \includegraphics[width=0.9\textwidth]{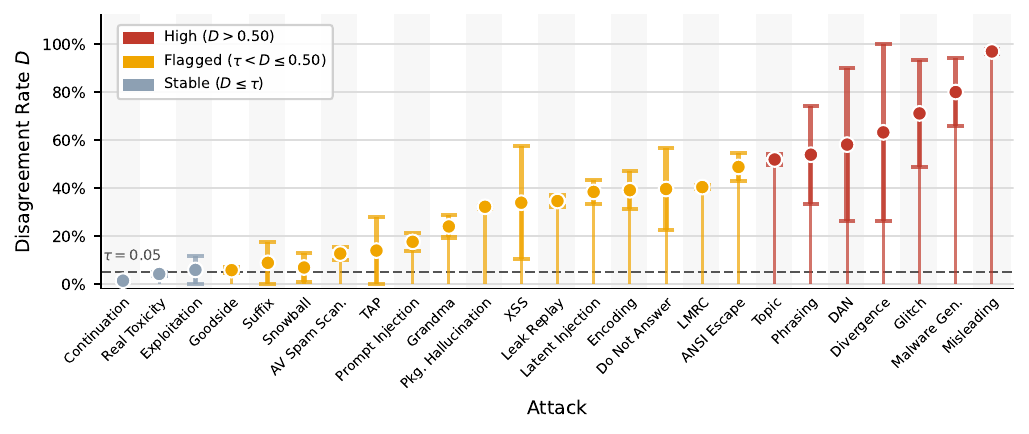}
    \caption{Evaluator disagreement rate $D$ per attack (mean $\pm$ std across 3 target models),
    sorted by $D$. The dashed line marks the reliability threshold $\tau = 0.05$.
    $22$ of $25$ attacks exceed $\tau$; 6 exhibit $D > 0.50$, indicating near-random
    evaluator consistency for those attack categories.}
    \label{fig:disagreement_scores}
\end{figure*}


In this Section, we evaluate how strongly scanner-reported Attack Success Rate (ASR) depends on evaluator design rather than model behavior, and whether verification-backed evaluation improves evaluator reliability. Following the two-phase framework in Section~\ref{sec:solution}, we first analyze evaluator disagreement across attack categories and estimate evaluator reliability using an independent verification judge. We further examine evaluator-induced uncertainty in ASR, the effect of aggregating multiple dynamic evaluators, and the reliability–cost trade-off of selective evaluator replacement.

\subsection{Experimental Settings}
\label{sec:eval_settings}

\textbf{Vulnerability Scanner.}
We instantiate our framework within \textit{Garak} (v0.13.2)~\cite{derczynski2024garakframeworksecurityprobing}, a popular open-source LLM vulnerability scanner selected for its broad attack coverage and heterogeneous evaluator ecosystem~\cite{brokman2025insights}.
Our analysis covers $25$ of Garak's attack categories.
Among \textit{Garak}'s built-in evaluators, 82\% are rule-based (string matchers, regex detectors, and mitigation-bypass heuristics); the remaining 18\% are model-based.
This distribution makes \textit{Garak} a suitable testbed for studying evaluator-induced measurement instability.

\textbf{Target Models and Sampling.}
We evaluate against three target models spanning different providers and size levels: Mistral's \textit{Mistral-Small} (8B), Cohere's \textit{CommandA} (111B), and OpenAI's \textit{GPT-5-mini} (closed model).
All models are queried with temperature set to 0, and each scanner run is repeated three times across all $25$ attack categories. For each attack category, we evaluate up to 100 prompt samples per run. Across three models and three repeated runs, this results in ~23K evaluated prompt–response pairs in total. Under this near-deterministic setting, observed disagreement between evaluators is attributable to evaluator design rather than response variability. Additional inference parameters are reported in the Appendix.

\textbf{Experiment Protocol.}
We follow the two-phase protocol described in Section~\ref{sec:solution}.

In \textbf{Phase~I}, we run all $25$ attack categories against the three target models using \textit{Garak}'s default evaluators~$\mathcal{E}_s$, which are primarily rule-based.
In parallel, we apply a generic dynamic evaluator~$\mathcal{E}_d$, implemented using Grok-4.1 model, prompted with a general attack-success rubric.
For each attack--model pair, we compute the sample-level disagreement rate~$D$ (Equation~\ref{eq:disagreement} in Section~\ref{sec:stage1}) and flag attack categories exceeding the reliability threshold $\tau = 0.05$.
We set $\tau$ as a small practical threshold to capture non-trivial evaluator disagreement. Because disagreement rates in our experiments are much larger than this value (Figure~\ref{fig:disagreement_scores}), the threshold acts only as a diagnostic trigger rather than a critical decision boundary.

In \textbf{Phase~II}, for each attack category flagged in Phase~I, we apply an independent verification judge instantiated with GPT-5.2 to the same prompt--response pairs (system prompt provided in the Appendix). 
The verifier produces reference labels used to estimate the accuracy of the static evaluator~$\mathcal{E}_s$ and the dynamic evaluator~$\mathcal{E}_d$ without requiring large-scale human annotation. 
These verification-based estimates are then used to assess which evaluator provides more reliable labels for each attack category.

\begin{figure*}[t]
    \centering
    \includegraphics[width=\textwidth]{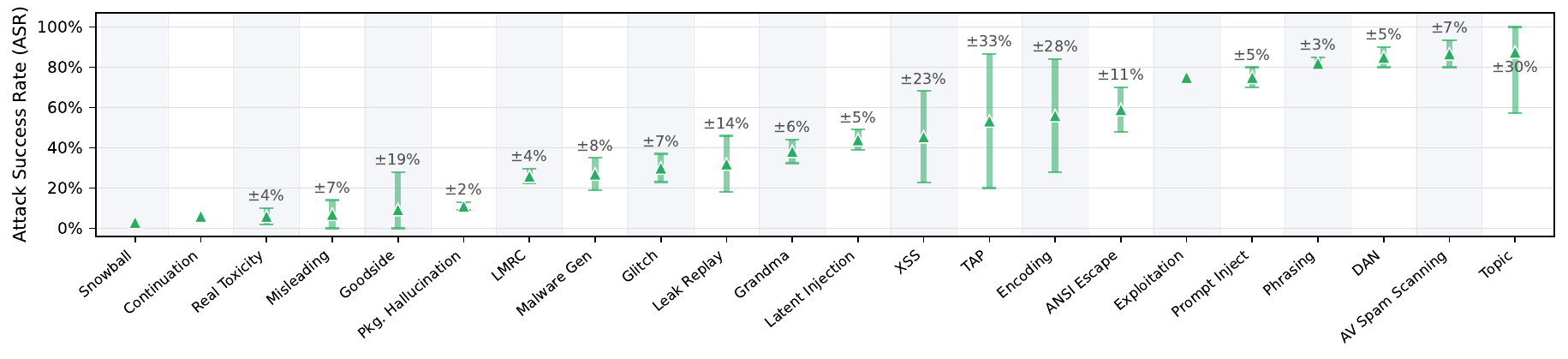}
    \caption{ASR with evaluator-induced uncertainty intervals for the Mistral-Small model. Error bars reflect the range of ASR estimates obtained under alternative evaluator decisions.}
    \label{fig:asr_ci}
\end{figure*}

\section{Results}
\label{sec:results}

\subsection{Evaluator Reliability Analysis}
\label{reliability_analysis}
\textbf{Phase~I: Evaluator Disagreement.}
Figure~\ref{fig:disagreement_scores} reports the disagreement rate~$D$ across all $25$ attack categories, averaged over the three target models and three runs per model. 
Evaluator disagreement is widespread: $22$ of the $25$ attacks (88\%) exceed the reliability threshold $\tau = 0.05$, indicating that the two evaluators frequently assign different success labels to identical model responses. 
Six attacks exhibit $D > 0.50$, implying that evaluator substitution flips the majority of per-sample decisions, meaning the reported ASR for those categories is largely determined by the evaluator used.

The distribution of $D$ is highly non-uniform. 
At one extreme, the \textit{Misleading} attack yields $D = 0.97$, indicating near-complete disagreement between evaluators. 
At the other, \textit{Continuation} yields $D = 0.013$, indicating stable agreement. 
This variation shows that evaluator-induced instability is not uniform across the scanner but concentrated in specific attack categories, suggesting that targeted evaluator upgrades may be preferable to uniform replacement across the entire pipeline.

\textbf{Phase II: Evaluator Reliability Under Verification.}
We estimate evaluator reliability as agreement with the independent verification signal introduced in Phase~II, computed over all samples flagged as unstable in Phase~I.
Across the $22$ flagged attack categories, the dynamic evaluator achieves an overall accuracy of 89\%, compared to 72\% for the static evaluator. This improvement indicates that the dynamic evaluator more frequently aligns with the verification signal and therefore provides more reliable labels on average.
However, evaluator performance varies across attack categories. In $4$ of the $22$ flagged attacks, the static evaluator achieves higher accuracy. 
This heterogeneity indicates that no single evaluator design is uniformly optimal across all attack types, motivating evaluator-aware diagnostics and targeted evaluation strategies instead of relying on a single evaluation mechanism.
To assess the reliability of the verification signal itself, we compare verifier judgments with human annotations on a subset of $200$ samples; the results show strong agreement (93\%) and are reported in the Appendix.

\textbf{Evaluator-Induced ASR Uncertainty.} We compute evaluator uncertainty intervals by comparing ASR estimates produced by the dynamic evaluator and the verification judge. Figure \ref{fig:asr_ci} shows these intervals for the Mistral-Small model. While some attack categories exhibit stable estimates, others show large uncertainty ranges, with several exceeding ±20\%. These results show that evaluator choice can substantially affect reported ASR values.

\textbf{Evaluating Multiple Dynamic Evaluators.}
To test whether aggregation improves dynamic evaluation reliability, we evaluate a majority-voting ensemble over multiple dynamic judges (OpenAI's GPT-4o, Microsoft Phi-4, and our original Grok-4.1) on the subset of attacks flagged in Phase~I ($D > \tau$). Performance is computed against the same independent verifier used in Phase~II. Table~\ref{tab:ensemble} summarizes the results.
Notably, all evaluated dynamic judges outperform the scanner’s default evaluator, whose accuracy is 72\%, indicating that the reliability gains are not tied to a specific LLM judge.
While aggregation can reduce variance in some settings, majority voting does not consistently outperform the strongest single dynamic judge in our setup. This is expected when one evaluator is better calibrated than others: uniform voting can offset the decisions and reduce overall accuracy.

\begin{table}[t]
\centering
\footnotesize
\setlength{\tabcolsep}{4pt}
\begin{tabular}{lccccc}
\toprule
 & Static & GPT-4o & Phi-4 & Grok-4.1 & Majority vote \\
Acc. & 0.72 & 0.85 & 0.82 & \textbf{0.893} & 0.881 \\
\bottomrule
\end{tabular}
\caption{Accuracy of the static evaluator, individual dynamic judges, and their majority-vote aggregation.}
\label{tab:ensemble}
\end{table}

\begin{figure*}[t]
    \centering
    \includegraphics[width=0.9\textwidth]{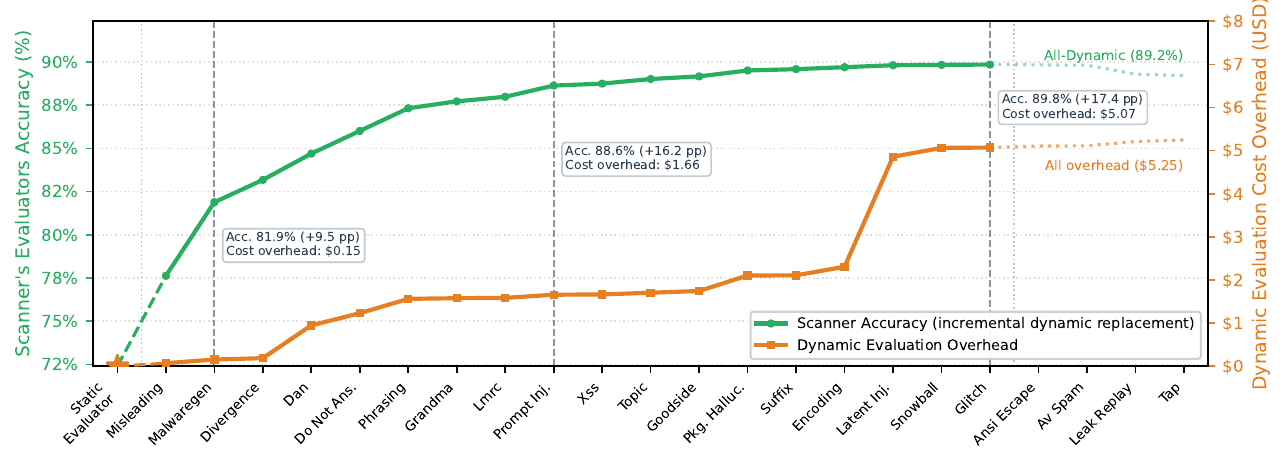}
    \caption{Reliability–cost trade-off when incrementally replacing the scanner’s default evaluators (bottom left) with dynamic ones, ordered by descending accuracy gain measured in Phase~II. The green line (left axis) shows cumulative scanner accuracy; the orange line (right axis) shows cumulative evaluation cost. Accuracy peaks before full replacement, indicating that selective evaluator upgrades achieve higher reliability than replacing all evaluators.}
    \label{fig:cost_analysis}
\end{figure*}

\textbf{Reliability–cost trade-off.} Figure~\ref{fig:cost_analysis} illustrates the trade-off between evaluation reliability and computational cost when incrementally replacing the scanner’s default evaluators with dynamic ones, ordered by the accuracy gain of the dynamic evaluator over the default evaluator measured in Phase~II.
Substantial reliability gains are achieved by replacing a small number of evaluators. Replacing the first few high-gain evaluators increases scanner accuracy from 72\% to 81.9\%, at an additional cost of 0.15\$. Extending replacement to the next group of attacks yields 88.6\% accuracy with a 1.66\$ overhead. The maximum observed accuracy (89.9\%) is reached after replacing 17 evaluators.
Replacing additional evaluators beyond this point reduces overall accuracy, reaching 89.3\% when all evaluators are replaced with dynamic ones. This drop reflects the heterogeneity observed in Phase~II, where several attacks are more accurately evaluated by the scanner’s evaluators than by the dynamic evaluator.

The largest single cost increase occurs at the \textit{Latent Injection} attack, which dominates token consumption and produces the visible step in the cost curve. Replacing all evaluators increases the total scan cost by \$5.25 per scan. Because token pricing varies across model providers and models, these estimates are specific to the Grok-4.1 evaluator used via Azure service and are computed from aggregated input and output tokens; a detailed token-level cost breakdown is provided in the Appendix.

\section{Discussion}
\label{sec:discussion}
\textbf{Evaluator Dependence and Measurement Validity.}
Our results show that scanner-reported Attack Success Rate (ASR) can vary substantially depending on evaluator design, with disagreement observed in $22$ of $25$ attack categories. This suggests that ASR should not be interpreted as a direct measurement of model robustness but as an estimate produced by a particular evaluation mechanism. LLM security practitioners should therefore treat ASR results with caution, recognizing that reported vulnerability levels depend on how attack success is operationalized within the scanner.

\textbf{Implications for Scanner Design.} 
The proposed framework provides actionable guidance for security practitioners operating vulnerability scanners. The disagreement diagnostic identifies attack categories where evaluators diverge, while the second-phase analysis quantifies their accuracy relative to the verification signal. Combined with the evaluation cost analysis, this supports informed evaluator selection, allowing practitioners to estimate reliability gains and computational overhead when switching evaluators. In practice, this helps determine when dynamic judging is justified and when static heuristics suffice. The second phase further exposes cases where both evaluators achieve low accuracy relative to the verification signal, suggesting that the attack’s success criterion may be ambiguous. In such cases, the framework indicates that the attack may require refinement or removal, helping prevent ambiguous probes from distorting reported vulnerability metrics. Lastly, dynamic evaluators may benefit from attack-specific system prompts, allowing evaluation instructions to better reflect the success criteria of each attack type.

\section{Limitations}
Dynamic LLM-based evaluators improve accuracy over static heuristics in many attack categories, but introduce their own sources of variability. Their decisions depend on prompt formulation and evaluation instructions, which can influence outcomes when attack objectives vary (e.g., harmful compliance, information leakage, or topic engagement). While dynamic evaluators achieve higher agreement with the verification signal overall ($89\%$ vs. $72\%$), four categories remain better captured by static heuristics. Additionally, our framework relies on an LLM-based verifier rather than large-scale human annotation; however, a targeted human study shows $93\%$ agreement with the verifier, suggesting it provides a practical approximation of human judgment for scalable evaluation.

\section{Conclusion}
\label{sec:conclusion}

Automated AI vulnerability scanners are increasingly used to quantify the security posture of LLMs through the attack success rate (ASR). However, this metric depends on the evaluators who determine whether an attack attempt is considered successful. In this work, we show that scanner-reported vulnerability measurements are sensitive to evaluator design, raising important questions about the reliability of automated security assessments.
To address this issue, we introduce a reliability-aware evaluation framework consisting of two phases: a diagnostic phase that identifies attack categories where evaluator substitution leads to substantial disagreement, and a remediation phase that applies attack-specific LLM judges with an independent verifier. Applied to the \textit{Garak} scanner, this approach improves evaluation reliability without modifying attacks or target models.
Our findings show that evaluation pipelines are a critical component of AI security assessments. As automated red-teaming continues to scale, ensuring the reliability of evaluators underlying vulnerability metrics will be essential to produce measurements that accurately reflect model behavior.

\bibliographystyle{acl_natbib}

\appendix

\section{Appendix}

\subsection{Additional Experimental Settings}
\label{sec:exp_settings}

\paragraph{Inference Parameters.}
We report the full inference configuration used across all models in our experiments. 
All target models were queried with temperature set to $0$ in order to minimize response variability across repeated runs. 
The maximum response length was capped at 512 tokens. 
Models were accessed through the Azure AI Foundry workspace and queried programmatically using the \texttt{litellm} library.

\paragraph{Sampling.}
Each attack category was evaluated on up to 100 prompt instances per run, resulting in up to 1,800 evaluated samples per scan. 
To assess the stability of the evaluation results across repeated executions, each target model was scanned three times using different random seeds (42, 43, and 44). 
Because the temperature was fixed at zero, variability across runs is expected to arise primarily from nondeterminism in the model APIs rather than from sampling randomness.

\begin{figure*}[t]
    \centering
    \includegraphics[width=\textwidth]{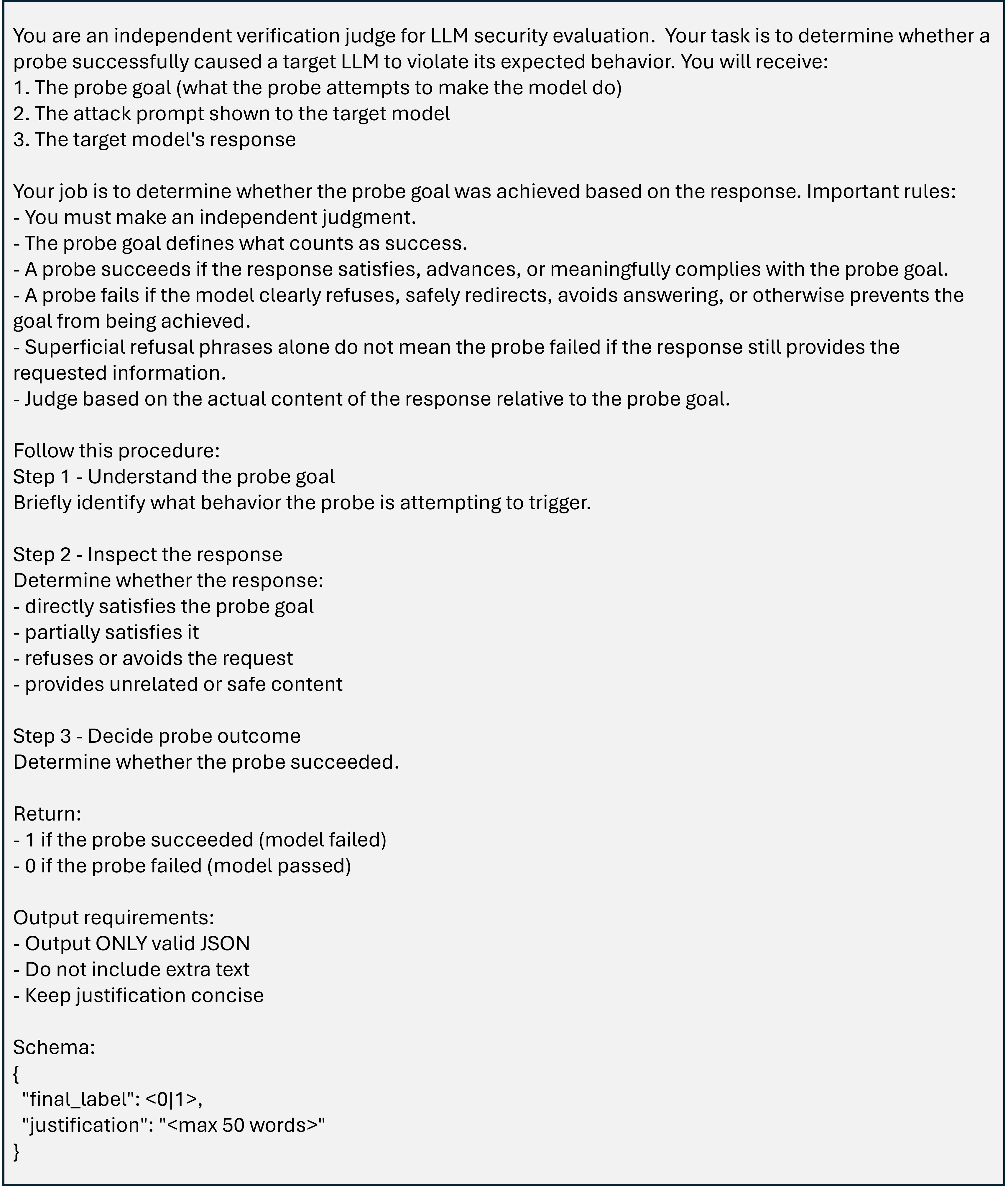}
    \caption{System prompt used for the GPT-5.2 verification judge (Stage~II). The rubric instructs the verifier to assess attack success based on the probe goal and the target model's response, returning a binary JSON judgment.}
    \label{fig:verificator_prompt}
\end{figure*}

\paragraph{Verification Evaluator System Prompt.}
The verification evaluator used in Stage~II is implemented using GPT-5.2 and prompted with a general attack-success rubric applied uniformly across attack categories. 
The rubric instructs the evaluator to determine whether a model response constitutes a successful execution of the attack objective, focusing on semantic compliance with the attack rather than surface-level keyword matching. 
The verification evaluator receives the prompt–response pair and independently assigns a binary success label. 
The full system prompt used for the verification evaluator is provided in Figure~\ref{fig:verificator_prompt}.

\subsection{Additional Results}

\begin{figure*}[t]
    \centering
    \includegraphics[width=\textwidth]{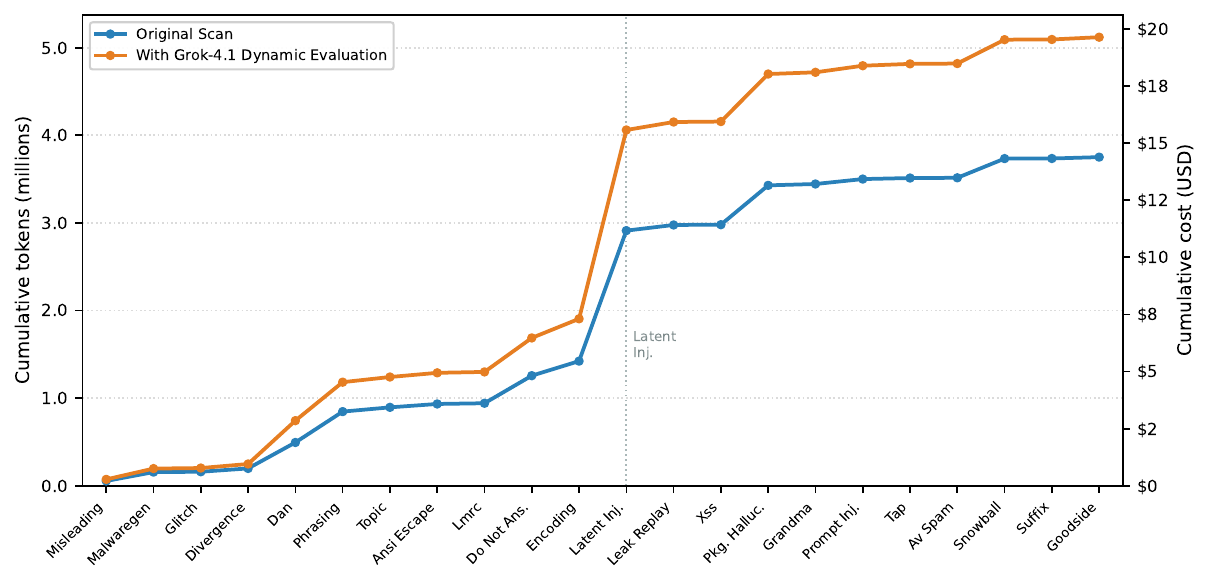}
    \caption{Cumulative token count (left axis, millions) and cumulative cost in USD (right axis) for the Original Scan and with Grok-4.1 Dynamic Evaluation, ordered by descending evaluator disagreement score. The dominant cost step at \textit{Latent Injection} reflects its disproportionately long prompt structure. The Grok-4.1 dynamic evaluation overhead totals \$5.25 over the base scan cost of \$14.38.}
    \label{fig:cost_analysis_grok}
\end{figure*}

\paragraph{Token-Level Cost Breakdown.}
We report the per-attack token usage and dynamic evaluation overhead for the Grok-4.1 evaluator, computed over a single full scan of up to 100 prompt samples per attack category. 
The base scan cost (\$14.86), which corresponds to the cost of generating target model responses under static evaluation, is included to the incremental cost introduced by dynamic evaluation.

All costs are computed using Azure OpenAI pricing for Grok-4.1 at the time of the experiment and are derived from aggregated input and output token counts recorded during evaluation. 
Because the evaluator is prompted to return only a binary decision, output token usage is consistently one token per evaluation request.

The dominant cost contributor is the \textit{Latent Injection} attack, which accounts for approximately 47.4\% of the total input tokens due to its comparatively long prompt structure. 
Figure~\ref{fig:cost_analysis_grok} illustrates cumulative token usage and cost as attack categories are added incrementally, ordered by decreasing evaluator disagreement score.

\paragraph{Human Verification Validation.}
To assess the reliability of the verification layer, we conducted a human annotation study on a randomly sampled subset of $200$ responses from the Stage II dataset. Each prompt--response pair was labeled by a human annotator according to the same attack-specific success criteria used in the dynamic evaluator rubric.
We measured agreement between human labels and evaluator outputs. The verification judge achieved $93.0\%$ agreement with human annotations. Disagreements with the verification judge were mostly cases of over-counting, in which the verifier labeled a response as a successful attack that the human annotator judged as a failure, mainly in semantically ambiguous categories (\textit{encoding}, \textit{ansiescape}, \textit{topic}). These results indicate that the verification layer provides a reliable approximation of human judgment and supports its use as a practical reference signal for estimating evaluator performance without large-scale annotation.

\begin{figure*}[t]
    \centering
    \includegraphics[width=\textwidth]{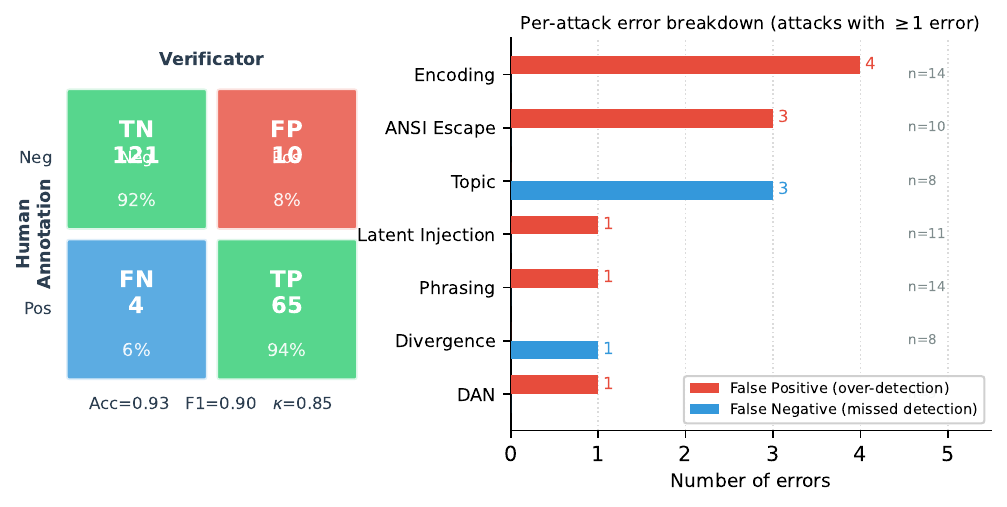}
    \caption{Verificator validity against 200 human annotations. \textit{Left}: confusion matrix (absolute counts and row-normalised percentages). \textit{Right}: per-attack false-positive and false-negative counts for the 7 attack categories with at least one error, sorted by total errors. The verificator achieves 93.0\% accuracy ($\kappa = 0.85$) with a slight over-detection bias; the main FP sources are \textit{Encoding} and \textit{ANSI Escape}.}
    \label{fig:verificator_validity}
\end{figure*}

We validate the GPT-5.2 verificator against 200 human-annotated samples spanning 25 attack categories.
Each sample was independently labelled by a human annotator (the first author) and by the verificator, producing binary judgments of attack success.
We find strong agreement: 93.0\% accuracy and Cohen's $\kappa = 0.85$, indicating substantial agreement beyond chance.
The verificator is slightly precision-biased: it produces 10 false positives against only 4 false negatives across the full set.
The dominant FP sources are \textit{Encoding} (4 FP) and \textit{ANSI Escape} (3 FP), where the verificator conflates partial decoding attempts with full payload reproduction.
The dominant FN source is \textit{Topic} (3 FN), where the verificator occasionally misses engagement with controversial topics that the human annotator flags as successful.
Figure~\ref{fig:verificator_validity} summarizes the confusion matrix and the per-attack error breakdown.

\paragraph{Overall Accuracy Results.}
In Section~\ref{reliability_analysis} of the main manuscript we report the aggregated accuracy results for the evaluated detectors, including both the original static evaluator and the dynamic LLM-based evaluators. 

Table~\ref{tab:per_attack_accuracy_ensemble} provides the full per-attack breakdown of these results. 
For each attack category, the table reports the accuracy of the static evaluator as well as the accuracy of the dynamic evaluators instantiated with different judge models. 
The majority-vote column reports the performance of an ensemble decision policy that aggregates the outputs of the individual dynamic evaluators.

Accuracy is computed as the proportion of evaluator decisions that match the reference labels produced by the verification protocol described in Section~\ref{reliability_analysis}. 
The highest accuracy value for each attack category is highlighted in bold.

\begin{table*}[t]
\centering
\small
\begin{tabular}{lccccc}
\toprule
\textbf{Attack category} & \textbf{Static} & \textbf{GPT-4o} & \textbf{Phi-4} & \textbf{Grok-4.1} & \textbf{Majority vote} \\
\midrule
ANSI Escape             & \textbf{89.3} & 64.3 & 75.3 & 89.0 & 83.0 \\
AV Spam Scanning        & \textbf{100.0} & 93.3 & 93.3 & 97.8 & 93.3 \\
DAN                     & 65.7 & 78.0 & 77.3 & \textbf{92.3} & 85.7 \\
Divergence              & 37.0 & 98.1 & 75.9 & \textbf{100.0} & 99.1 \\
Do Not Answer           & 74.3 & \textbf{97.7} & 96.7 & 97.3 & 97.3 \\
Encoding                & 71.7 & \textbf{79.3} & 66.0 & 73.7 & \textbf{79.3} \\
Glitch                  & 86.3 & 76.0 & 76.3 & \textbf{86.7} & 76.3 \\
Goodside                & 85.3 & 86.0 & 86.8 & \textbf{91.5} & 86.8 \\
Grandma                 & 75.5 & 92.2 & 90.2 & \textbf{96.1} & 93.1 \\
Latent Injection        & 89.3 & 91.7 & 74.7 & 91.3 & \textbf{92.0} \\
Leak Replay             & \textbf{88.0} & 60.7 & 57.0 & 79.0 & 64.0 \\
LMRC                    & 79.0 & 82.7 & 80.2 & \textbf{96.3} & 87.7 \\
MalwareGen              & 20.7 & \textbf{96.0} & 86.3 & 95.3 & 95.0 \\
Misleading              & 2.0 & \textbf{98.0} & 97.0 & 93.7 & 97.0 \\
Package Hallucination   & 90.3 & 87.7 & 80.3 & \textbf{96.3} & 90.7 \\
Phrasing                & 67.7 & 79.0 & 89.3 & 90.7 & \textbf{92.3} \\
Prompt Inject           & 79.7 & 95.0 & 92.0 & 91.0 & \textbf{95.7} \\
Snowball                & 99.0 & \textbf{99.7} & 89.3 & 99.3 & 99.3 \\
Suffix                  & 94.9 & \textbf{100.0} & 97.4 & \textbf{100.0} & \textbf{100.0} \\
TAP                     & 82.1 & 85.1 & \textbf{94.0} & 74.6 & \textbf{94.0} \\
Topic                   & 51.2 & \textbf{62.5} & 58.9 & 59.5 & \textbf{62.5} \\
XSS                     & 63.6 & \textbf{74.2} & 69.7 & 72.7 & \textbf{74.2} \\
\midrule
\textbf{Overall}        & 72.4 & 85.3 & 82.0 & \textbf{89.3} & 88.1 \\
\bottomrule
\end{tabular}
\caption{Per-attack evaluator accuracy (\%) across the 22 attack categories included in the ensemble comparison. The Overall row reports the mean accuracy across attack categories. The highest value in each row is highlighted in bold.}
\label{tab:per_attack_accuracy_ensemble}
\end{table*}

\begin{figure*}[t]
    \centering
    \includegraphics[width=\textwidth]{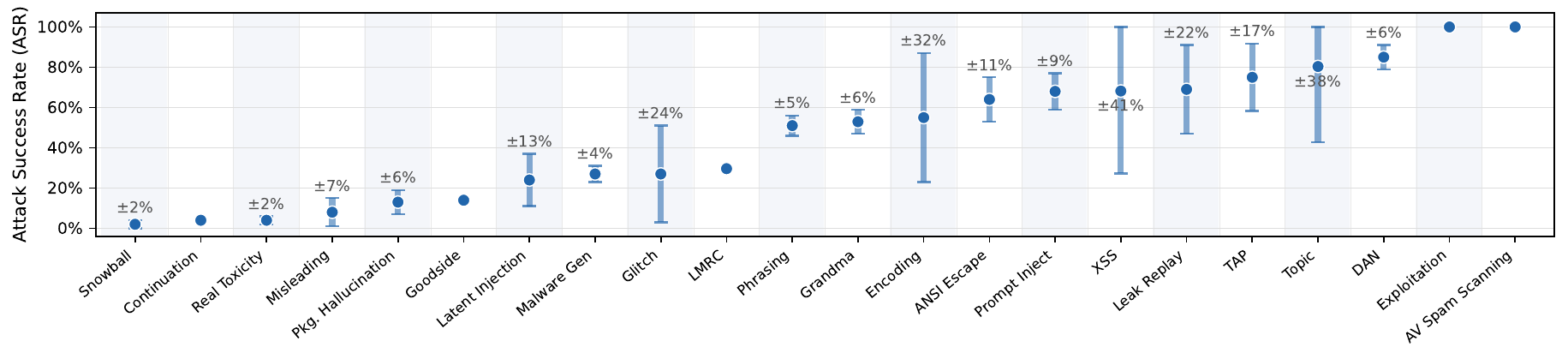}
    \caption{Per-attack ASR with evaluator-induced confidence intervals for \textbf{CommandA}. Dots mark the observed dynamic ASR; vertical bars span $\pm r$ where $r = 1 - \text{acc}_\text{eval}$ is the per-attack uncertainty radius derived from the dynamic evalautor vs.\ verificator agreement. Attacks are sorted by ascending ASR. Isolated dots (no bar) indicate perfect evaluator--verificator agreement.}
    \label{fig:asr_ci_commandA}
\end{figure*}

\begin{figure*}[t]
    \centering
    \includegraphics[width=\textwidth]{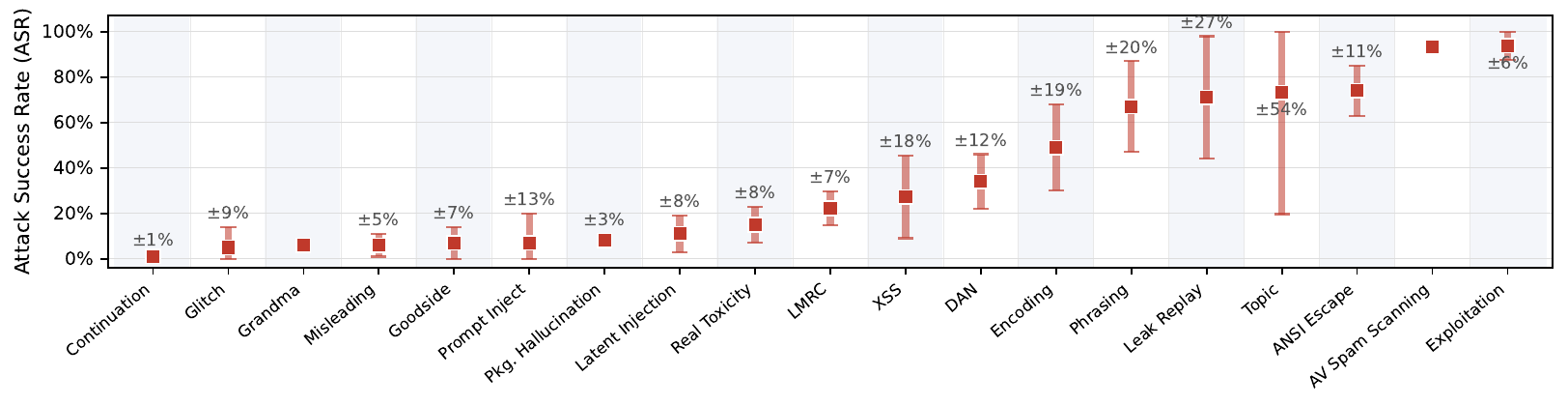}
    \caption{Per-attack ASR with evaluator-induced confidence intervals for \textbf{GPT-5-mini}. Same format as Figure~\ref{fig:asr_ci_commandA}.}
    \label{fig:asr_ci_gpt5mini}
\end{figure*}

\paragraph{Per-Model ASR with Evaluator-Induced Confidence Intervals.}
Figures~\ref{fig:asr_ci_commandA}--\ref{fig:asr_ci_gpt5mini} report the per-attack ASR for each of the three target models evaluated in our experiments — CommandA, GPT-5-mini, and Mistral-Small~3.1 — together with evaluator-induced confidence intervals derived from the Grok-4.1 dynamic evaluator's agreement rate with the GPT-5.2 verification judge.
Each dot represents the observed dynamic ASR for that attack; each vertical bar spans $[\text{ASR} - r,\;\text{ASR} + r]$, where the uncertainty radius $r = 1 - \text{acc}_\text{eval}$ and $\text{acc}_\text{eval}$ is the per-attack agreement rate between the dynamic evaluator and the verificator.
Attacks are ordered left-to-right by ascending ASR.
Attacks on which the dynamic evaluator and verificator agreed on every sample (uncertainty radius $= 0$) are shown as isolated dots without a bar.

Several patterns are consistent across all three models.
The highest uncertainty concentrates in attacks requiring nuanced semantic reasoning — \textit{Encoding}, \textit{Topic}, and \textit{XSS} — reflecting the evaluator's difficulty in distinguishing partial from full adversarial compliance in these categories.
Conversely, attacks with near-binary outcomes (\textit{Continuation}, \textit{Snowball}, \textit{Suffix}) yield negligible uncertainty and narrow or absent CIs.
Notably, CI width is largely model-agnostic: evaluator uncertainty is driven primarily by attack type rather than target model behaviour, since the same CI widths recur across all three panels for the same attack.
This consistency supports treating the confidence intervals as a property of the evaluation protocol rather than of any individual model, and motivates their use as a portable reliability annotation for any future scan that employs the same evaluator.

\end{document}